\documentclass{aa520}
\usepackage{graphics,natbib}
\bibpunct{(}{)}{;}{a}{}{,}

\begin{document}

\title{Multiple minor mergers: formation of elliptical galaxies\\ and constraints for the growth of spiral disks}
\author{Fr\'ed\'eric Bournaud \inst{1}, Chanda J. Jog \inst{2}, and Fran\c{c}oise Combes \inst{3}}\offprints{F. Bournaud \email{frederic.bournaud@cea.fr}}
\institute{Laboratoire AIM, CEA-Saclay DSM/DAPNIA/SAp -- CNRS -- Universit\'e Paris Diderot, 91191 Gif-sur-Yvette, France
\and
Department of Physics, Indian Institute of Science, Bangalore 560012, India
\and
Observatoire de Paris, LERMA, 61 Av. de l'Observatoire, F-75014, Paris, France}
\date{Received; accepted}

\abstract{Multiple, sequential mergers are unavoidable in the hierarchical build-up picture of galaxies, in particular for the minor mergers that are frequent and highly likely to have occured several times for most present-day galaxies. However, the effect of repeated minor mergers on galactic structure and evolution has not been studied systematically so far. We present a numerical study of multiple, subsequent, minor galaxy mergers, with various mass ratios ranging from 4:1 to 50:1. The N-body simulations include gas dynamics and star formation. We study the morphological and kinematical properties of the remnants, and show that several so-called ``minor'' mergers can lead to the formation of elliptical-like galaxies that have global morphological and kinematical properties similar to that observed in real elliptical galaxies. The properties of these systems are compared with those of elliptical galaxies produced by the standard scenario of one single major merger. We thus show that repeated minor mergers can theoretically form elliptical galaxies without major mergers, and can be more frequent than major mergers, in particular at moderate redshift. This process must then have formed some elliptical galaxies seen today, and could in particular explain the high boxiness of massive ellipticals, and some fundamental relations observed in ellipticals. In addition, because repeated minor mergers, even at high mass ratios, destroy disks into spheroids, these results indicate that spiral galaxies cannot have grown only by a succession of minor mergers.

}
\authorrunning{Bournaud et al.}
\titlerunning{Multiple minor mergers and the formation of elliptical galaxies}
\maketitle


\section{Introduction}

Major mergers between spiral galaxies of similar masses are known to form elliptical-like galaxies. The remnants of such violent events present an $r^{1/4}$ radial mass profile, as observed in elliptical galaxies \citep{dvc77}, and are mainly pressure-supported. This is the case for equal-mass mergers \citep[e.g.][]{BH91starb, barnes92}, but more generally for mass ratios ranging from 1:1 to 3:1 and even 4:1, studied by \citet{bendobarnes}, \citet{Cretton01}, \citet{naabburkert03} and in Bournaud, Jog \& Combes (2005b, hereafter paper~I). The merger remnant tends to be boxy for 1:1 mergers, while 3:1 and 4:1 mergers mostly result in disky elliptical galaxies -- that are not disk galaxies, but elliptical galaxies with a disky isophotal shape. An outer disk-like component appears in particular for mass ratios of 3:1 and higher \citep{naabtrujillo06}. It can be present even in 1:1 or 2:1 merger remnants but is far from being the main contribution to the total stellar mass. The disk component becomes more massive with increasing mass ratios. Beyond 4:1, mergers do not form elliptical galaxies any longer, but early-type disk-dominated systems (Bournaud, Combes \& Jog 2004, and paper~I). They are usually called ``minor'' mergers, but we have suggested in these earlier works that one should distinguish between ``intermediate'' mergers (between 4:1 and around 10:1) that form S0-like galaxies\footnote{These systems have a disk-like profile, but a massive bulge and their support is dominated by pressure instead of rotation, unlike spirals, making them rather S0-like \citep{jogchitre02, bourn04merg}}, and the real ``minor'' mergers with ratios larger than 10:1, where the remnants can be classified as (disturbed) spiral galaxies (Quinn, Hernquist \& Fullagar 1993; Velazquez \& White 1999; Walker, Mihos \& Hernquist 1996).

Existing studies of multiple mergers have focused on nearly-simultaneous mergers of several galaxies of comparable masses \citep{WeilH94,WeilH96,bekki01}. This typically corresponds to the collapse of a compact group into one single giant elliptical galaxy. \citet{WeilH94,WeilH96} have shown that the elliptical galaxies produced by these events have some kinematical properties that differ from the remnants of pair mergers, and provide a better explanation for observed properties. \citet{bekki01} has studied the starbursts occurring when such events involve gas-rich late-type spirals, and the possible connection with ULIRGs. There is observational evidence that some galaxies are remnants of the collapse of compact groups \citep[e.g.,][]{borne2000} or at least of the simultaneous merging of more than 2 galaxies \citep{Taniguchi98}. However, these studies are limited to situations where the different galaxies merge nearly at the same time, and have comparable masses, so that the merger of only two of them would already have resulted in the formation of an elliptical galaxy. 

In this paper, we study a different process: the multiple, sequential mergers of the intermediate and minor types -- hereafter both called ``minor'' for simplicity -- i.e. the mass ratios that do not form elliptical-like galaxies after one single merging event. This corresponds to a given spiral galaxy that gradually grows by merging with smaller companions. The mergers are not assumed to occur at the same time, but one after the other, over a total timescale of a few Gyrs. We show that a few successive mergers with mass ratios between 5:1 and 10:1 lead to the gradual transformation of spiral galaxies into elliptical-like systems. Under the effects of the sequential mergers, spiral galaxies become earlier-type spirals, then lenticular S0-like systems, and finally spheroidal objects with global morphological and kinematical properties similar to observed elliptical galaxies. This new scenario for the formation of elliptical galaxies is compared with the result of the ``standard'' binary major merger scenario, although the differences are not striking. We also show that even the growth of a spiral merging with 50:1 dwarf companions also leads to the progressive destruction of the spiral disk and the evolution towards S0-like and more frequently elliptical-like galaxies, if the initial galaxy is assumed to significantly increase its mass by such mergers with dwarf companions. 
Repeated minor mergers occurring alone can then in theory be an efficient process to form a large number of elliptical galaxies. However, we discuss that in reality other processes can reduce this efficiency, in particular to maintain massive disks in spiral galaxies. At the same time, some elliptical galaxies can likely have formed via multiple minor mergers, in particular the boxy ones and/or those formed at low redshift when minor mergers largely dominate major ones.

Section~2 contains the details of N-body simulations and analysis of the results. In Sect.~3, we analyze the properties of the multiple merger remnants as a function of the mass ratios and the number of mergers, and study the differences with or similarities to the elliptical galaxies resulting from major mergers. The implications for galaxy evolution are studied in Sect.~4, and Sect.~5 contains a brief summary of the results of this paper.


\section{N-body simulations of multiple galaxy mergers}

\subsection{Code description}

We have used the N-body FFT code of \citet{BC03}. The gravitational field is computed with a resolution of 400pc. We used $4 \times 10^6$ particles for the most massive galaxy. The number of particles used for the other galaxy is proportional to its mass. Star formation and time-dependent stellar mass-loss are implemented as described in \citet{BC02}.

The star formation rate is computed according to the generalized Schmidt law \citep{schmidt59}: the local star formation rate is assumed to be proportional to ${\mu_g}^{\beta}$, where $\mu_g$ is the local two-dimensional density of gas. We chose $\beta=1.5$, as suggested by the observational results of \citet{kennicutt98} \citep[see also][]{boissier06}. The dissipative dynamics of the ISM has been accounted for by the sticky-particles scheme described in Appendix~A of \citet{BC02}. In this paper we employ elasticity parameters $\beta_t$=$\beta_r$=0.7. 

\subsection{Physical model for colliding galaxies}

Each galaxy is initially made up of a stellar and gaseous disk, a spherical bulge and a spherical dark halo. The visible mass of the main (target) galaxy is $2\times 10^{11}$ M$_{\sun}$. Its stellar disk is a \citet{toomre64} disk of radial scalelength 5~kpc, truncated at 15~kpc. Gas is distributed in a Toomre disk of scale-length 15~kpc and radius 30~kpc. The bulge and dark halos are Plummer spheres of radial scalelengths 3~kpc and 30~kpc respectively. The dark halo is truncated at 70~kpc. The bulge-to-total mass ratio is 0.17 (bulge-to-disk: 0.2), and the dark-to-visible mass ratio inside the stellar disk radius is 0.7. This corresponds to a stellar disk of $14\times 10^{10}$ M$_{\sun}$, gas disk of $2.5\times 10^{10}$ M$_{\sun}$, a bulge of $3.3\times 10^{10}$ M$_{\sun}$, and a dark halo of $11.7\times 10^{11}$ M$_{\sun}$ (up to $r=70$~kpc). The number of particles is $1.3\times 10^6$ for stars, $0.93\times 10^6$ for gas, and $1.76\times 10^6$ for dark matter.

The initial velocities of particles for each components are computed as in \citet{BC03}. The initial value of the Toomre parameter is $Q=1.7$ over the whole disk. The gas mass fraction in the disk is 15\%. In this paper, we wish to compare successive minor mergers to binary major mergers formed under the same conditions, so we do not vary the gas fraction, except for the 50:1 mergers presented later in Sect.~\ref{412}.

The companion galaxies have a radial distribution of matter scaled by the square root of their stellar mass. For instance a 4:1 companion has a disk that is half the size, as is the bulge, etc.. Gas mass fraction, bulge-to-disk and halo-to-disk mass ratios are kept constant to avoid a too large number of free parameters. The free parameters are then:

\begin{itemize}
\item the number of merging companions
\item the mass ratio of each companion
\item the orbital parameters of each companion
\end{itemize}

\subsection{Orbital parameters}

The orbital parameters of a merging pair of galaxies have been described e.g. by \citet{duc2000}. These authors defined two angles $\theta$ and $\Phi$ for each galactic disk. $\theta$ is the inclination of the orbital plane vs the primary disk plane ($\theta=0$ for a prograde encounter in the disk plane, $\theta=180$ for a retrograde encounter in the disk plane). $\Phi$ is the azimuthal position of the secondary disk spin axis. We have run series of simulations with $\Phi$ fixed to 30 degrees and a value of $\theta$ of either 30 or 180+30 degrees, for each companion (i.e. inclinations of 30 degrees but orbits either prograde or retrograde). The first encounter is prograde with a corotating companion, the second one is retrograde, with a counter-rotating companion, etc.. We fixed the encounter velocity to 150~km~s$^{-1}$ and impact parameter to 45~kpc, both computed at an infinite distance, neglecting the action of dynamical friction before the beginning of the simulation\footnote{These values should be compared to the stellar disk radius of 15~kpc and circular velocity 180~km~s$^{-1}$ at the edge of this disk, to be applied to smaller/larger galaxies}. Since the velocity at an infinite distance is larger than zero, the initial orbits are not bound but hyperbolic with a total energy larger than zero, and dynamical friction dissipates the relative kinetic of the galaxies, eventually leading to a merger.

The choice of an inclination of 30 degrees is made because it is near the statistically average value of the inclination angle in spherical geometry. Assuming an isotropic distribution of the initial orbital planes of galaxies, the probability of an inclination $i$ is $p(i) \propto \cos(i)$, and the average inclination $ < i > = \int \; i p(i) \; d i \simeq 30\deg$. This average choice is then representative of orbits that are neither coplanar nor polar. The inclination has not been varied in this paper, except for sequences of 10:1 mergers as described in Sect.~\ref{411}. It is known from paper~I and other studies of major mergers that the properties of a binary merger remnant depend more on the mass ratio than the inclination and other orbital parameters; the test in Sect.~\ref{411} confirms that the result of a minor merger sequence depends on the total merged mass more than the orbits, even if changing the orbit does result in some variations of the final properties.

In the reference case of one binary equal-mass merger, choosing one orbit orientation (prograde or retrograde) may not be representative; we thus simulated the two cases and kept the average properties in the following study: we still call this ``run {\tt 1x1:1}'' for simplicity, and the properties of a merger remnants depend more on the mass ratio itself then the orbital parameters (e.g., paper~I and other references in the Introduction). Similarly, for the 3:1 cases, we ran two sequences (prograde-retograde-prograde and retrograde-prograde-retrograde) and kept the average results. For higher mass ratios, we discuss larger number of mergers so that starting with the first orbit as prograde or retrograde has less influence -- the influence of the orbit orientation is anyway smaller than the influence of the mass ratio itself (paper~I for binary mergers, see also Sect.~\ref{411} in this paper for multiple mergers).

\subsection{Temporal sequences of mergers}

The nomenclature for each run is as follows: 
{\tt NxM:1} indicates that the galaxy has undergone {\tt N} mergers with companions of mass $1/${\tt M} of the target galaxy mass.\\
We have simulated mass ratios {\tt M} of 1, 3, 5, 7, 10, 15, and {\tt N} varying from 1 to {\tt M} (by steps of 2 for {\tt M}=15). 

In sequences with mass ratios up to 10:1, the mergers are separated in time by $\Delta t = 800$~Myr. This means that each interloper will reach its pericenter at an instant computed to be $\Delta t$ after the pericenter of the previous ones. In studies of binary mergers over the same range of mass ratios (paper~I) we found that this timescale ensures that each merger is relatively relaxed\footnote{In particular, this 800~Myr interval {\emph{after}} the first pericenter ensures that the nuclei from the first merger have coalesced when the second interaction is at its pericenter.} before the following one occurs -- even if the merging/relaxation occurs somewhat more rapidly for 1:1 cases than 10:1 cases. A time interval of 2~Gyrs or more would ensure a more complete relaxation of the each merger before the next one occurs, but this would lead to total durations longer than the Hubble time. Actually, multiple mergers are rather expected to occur in groups or dense cosmological structures, with a total duration of the merger sequence of a few Gyrs and a somewhat short time interval between mergers.

As for 15:1 mass ratios, the injection of fifteen companions each 800~Myr would represent a total duration of 12~Gyr, which is highly unrealistic, in particular with the adopted mass and gas fraction of the initial target galaxy. We then chose to reduce the time interval between mergers to $\Delta t = 400$~Myr for this 15:1 mass ratio. The first merger is only partially relaxed when the second occurs, but the mergers are still subsequent, i.e. the nuclei of the first companion is merged when the second companion reaches the outer gas disk at $r=30$~kpc.

The time interval separating mergers is not varied in the present paper. We do not find strong differences between the 10:1 ($\Delta t = 800$) and 15:1 ($\Delta t = 400$) merger sequences in the following, suggesting that it has only a minor influence on the main properties of the remnant galaxies, at least as long as the mergers are sequential and not simultaneous. Detailed properties like the orbital structure could however vary with the time interval separating the merging events \citep[e.g.][]{athan05}; in particular, in the present paper we study subsequent mergers and do not compare to simultaneous mergers.

In each run, the final system is evolved 1.5~Gyr after the last merger, to ensure full relaxation before any analysis is performed. This way, the {\tt 5x10:1} simulation corresponds to five mergers of mass ratio 10:1, separated in time by 800~Myr between their respective pericenter passages, followed by a 1.5~Gyr isolated evolution to ensure the relaxation of the the final remnant before its properties are analysed. In the the {\tt 10x10:1} simulation, there are 9 intervals of 800~Myr between the mergers and a final 1.5~Gyr-long relaxation of the remnant.

We also ran a simulation with 50 dwarf companions of mass ratio 50:1, and initial gas mass fraction of 30\%. This simulation, where the mergers cannot really be subsequent, but overlap over time, is described later in Sect.~\ref{412}. We do not vary the orbital parameters beyond alternating prograde and retrograde orbits in the main part of the paper, but we will describe and discuss the influence of orbital parameters in {\tt 10x10:1} merger sequences in Sect.~\ref{411}.

\subsection{Analysis of the results}

Each remnant of a (sequence of) mergers is analyzed, once relaxed, to derive several physical quantities. We build projected column-density maps on which we define a ''25th- isophote'', its semi-major-axis being the $R_{25}$ radius in the following, commonly called the ''optical radius'' in observations. The conversion of mass density into luminosity is calibrated so that the initial target spiral galaxy, seen face-on, has a central surface magnitude $\mu$=21.7 for its disk component (not counting the additional luminosity from the bulge). This is to be consistent with the \citet{freeman70} relation, where this central magnitude is primarly observed in B-band, but we do not explicitly assume any wavelength in the analysis of our simulations. We assume that the mass-luminosity conversion factor does not vary over the duration of our simulations, because we start with 15\% of gas so that the new stars will represent at most 15\% of the mass\footnote{actually most new stars formed in mergers lie in the central kpc, not around $R_{25}$, which results in an even lower contamination at this radius}. Most stars are older than the merging events, hence have a reasonably constant mass-to-light ratio. The 25th isophote is found to include, on average, 82\% of the stellar mass of the merger remants in our simulations. Once this reference radius $R_{25}$ is defined, we can measure several fundamental parameters of the multiple merger remants:

\begin{itemize}
\item the {\emph{flattening}} $E$. We first derive the ``face-on'' projection, defined as that minimizing the apparent flattening of the 25th isophote. We then construct projected maps for ten ``edge-on'' projections, and measure the ellipticity $e(r)=10 \times (1-b/a)$ where the axis ratio $b/a$ is obtained from an ellipse-fitting model. We finally keep the flattening $E$ as the average value of $e(r)$ over the $[0.55R_{25};R_{25}]$ range for the different projections. This definition is similar to that used in Bournaud et~al. 2005b, and roughly equivalent to that used by \citet{naabburkert03}. 

\item the {\emph{Sersic index}} $n$ is determined as the best fit over the same $[0.55R_{25};R_{25}]$ radial range, averaged for three orthogonal projections (not restricted to ``edge-on'' projections as was the case when deriving the flattening $E$).

\item the {\emph{kinematical parameter}} $V/\sigma$. When building projected images we also compute the velocity distribution along each line-of-sight, and then derive the average velocity $V$ and velocity dispersion $\sigma$ for each one. We then define the $V/\sigma$ parameter as the column-density-weighted average (equivalent to luminosity-weighted assuming a constant mass-to-luminosity ratio) over the $[0.55R_{25};R_{25}]$ radial range, and the result is averaged over three orthogonal projections. The $V/\sigma$ parameter is not measured along a slit, but on all the pixels of a first momentum map within the chosen radial range, each pixel being weighted by its intensity in the associated density map. The choice to exclude the central regions ($r<0.55R_{25}$) from this kinematical analysis was made to ensure that a disk galaxy will have a $V/\sigma$ parameter significantly larger than 1. Indeed this lower bound is always larger than 5~kpc in our merger remnants, which is larger than the bulges of spirals and S0s (even the large and massive bulges of intermediate-mass merger remnants in Bournaud et al. 2004). This way the presence of a massive rotating component will result in a $V/\sigma > 1$, while a value of $V/\sigma < 1$ cannot relate to the bulge of a spiral galaxy, and indicates elliptical-like kinematics at large scales.

\item the {\emph{boxiness}} $a_4$. When measuring $E$ from various projections, we also compute the boxiness $a_4(r)$ as a function of the radius, and keep the average value $a_4$ over the range $[0.55R_{25};R_{25}]$. The choice of $0.55R_{25}$ as the lower bound ensures that, when a massive disk is present, we really measure its boxiness or diskiness, not that of the bulge (see paper~I).

\end{itemize}


\section{Formation of elliptical galaxies in successive mergers}

\subsection{First detailed example}

\begin{figure*}
\resizebox{11cm}{!}{\includegraphics{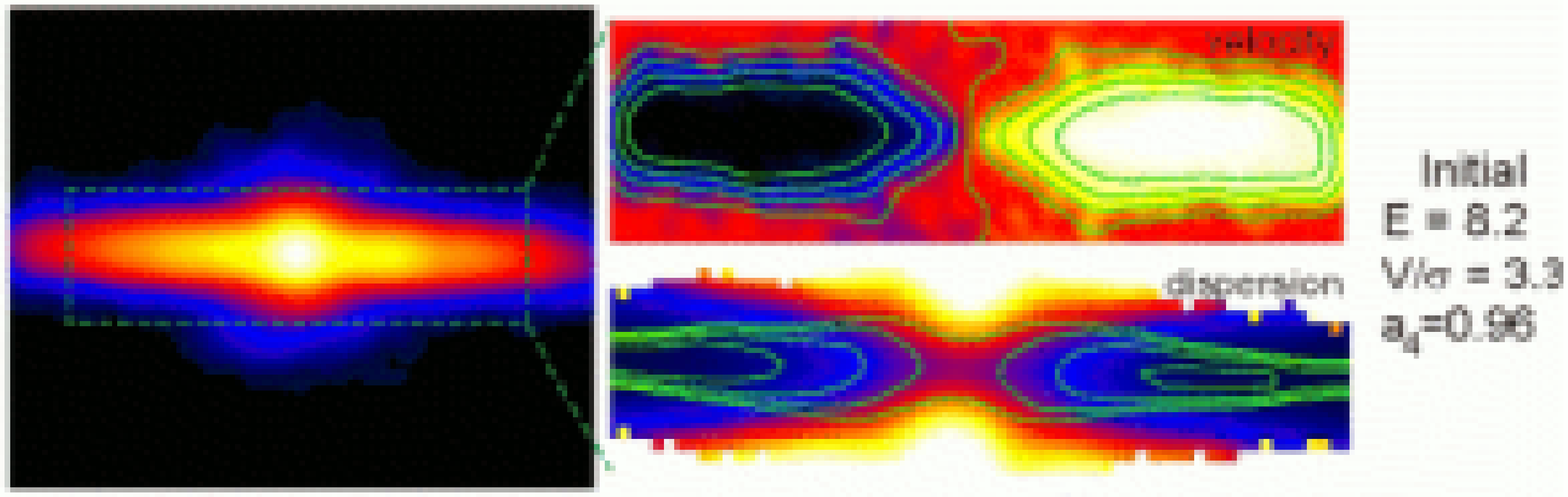}}\\
\resizebox{11cm}{!}{\includegraphics{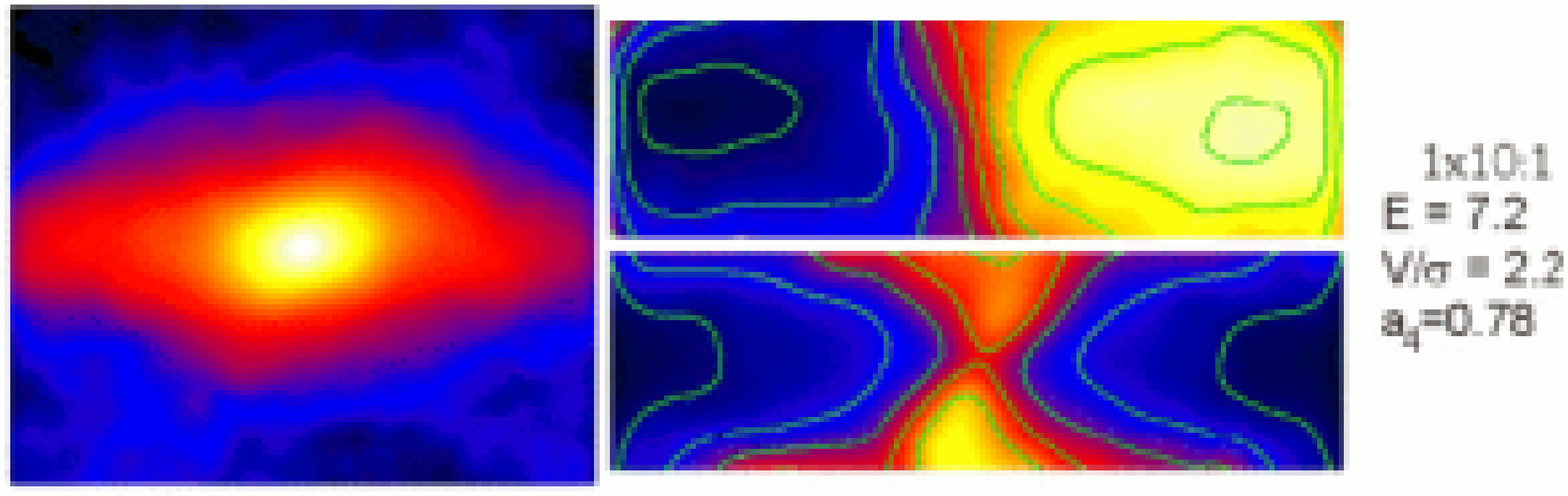}}\\
\resizebox{11cm}{!}{\includegraphics{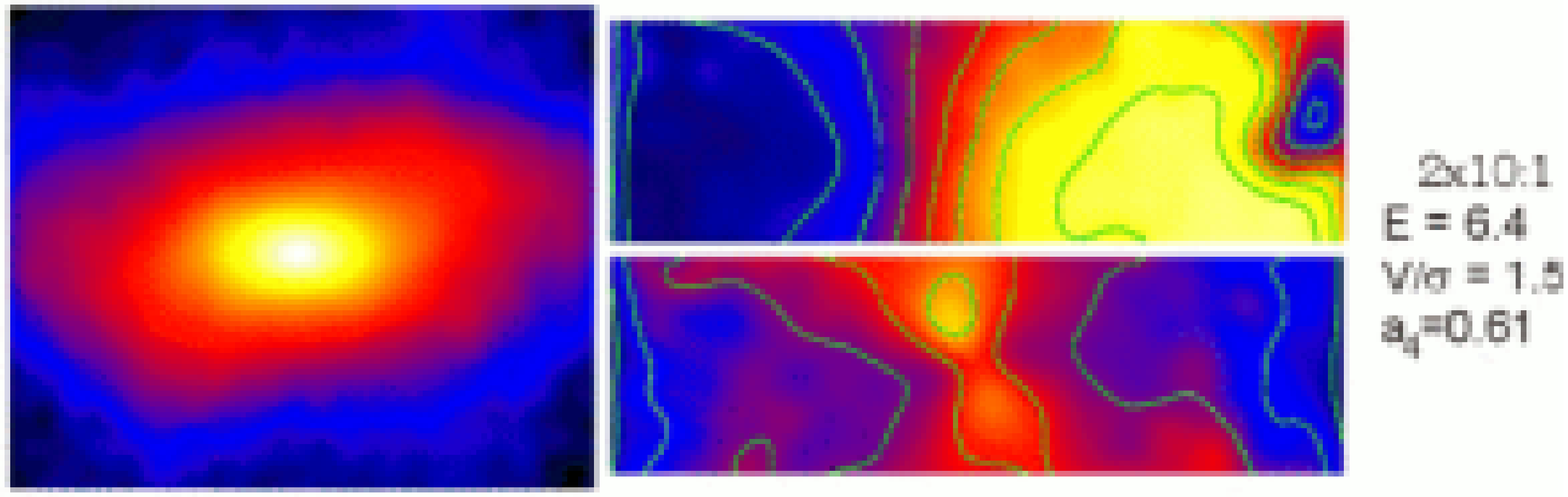}}\\
\resizebox{11cm}{!}{\includegraphics{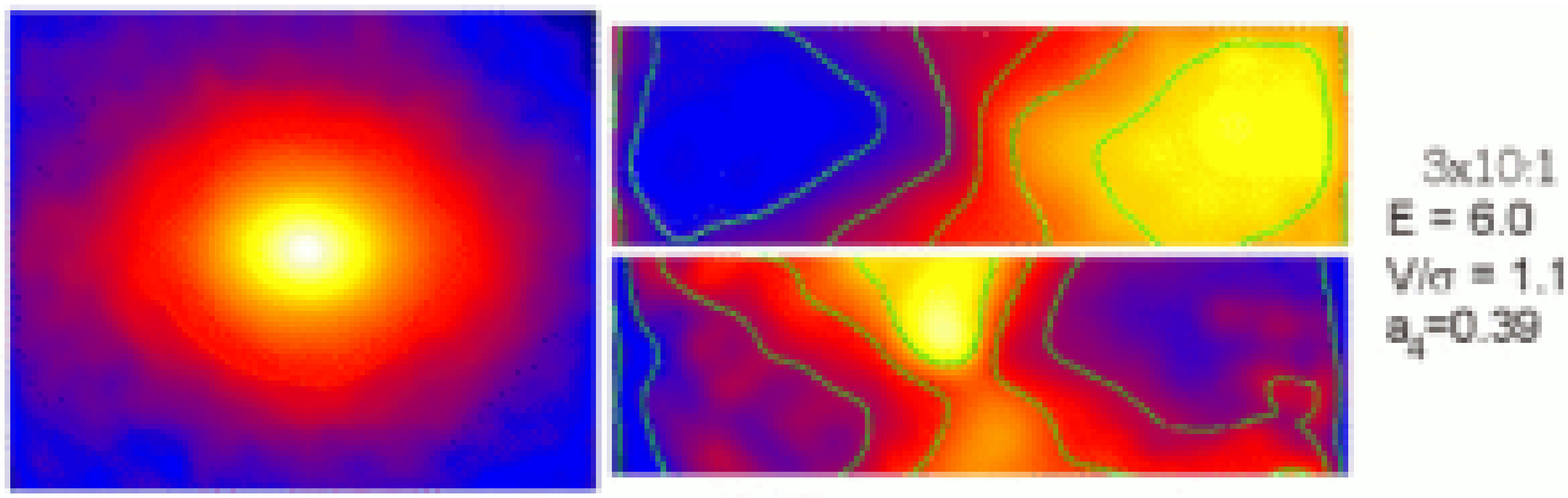}}\\
\resizebox{11cm}{!}{\includegraphics{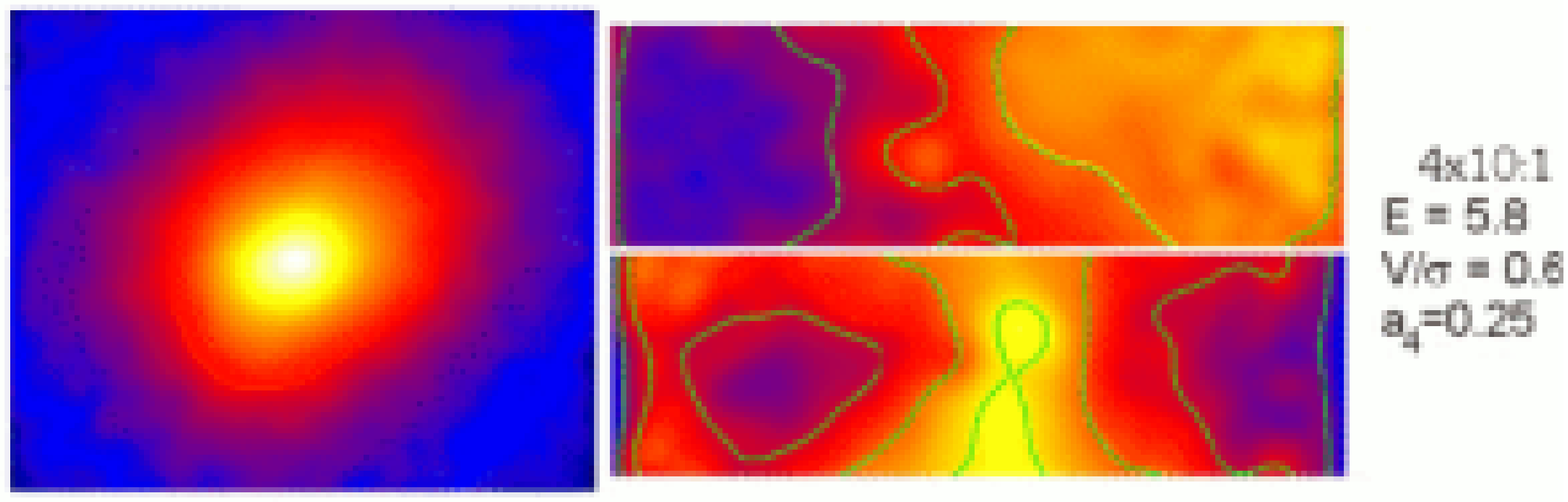}}\\
\caption{Comparison of the initial spiral galaxy (here seen after 1 billion year of isolated evolution) and the relaxed merger remnants of one to four 10:1 mergers. The density maps (left) are in logscale. The velocity and dispersion fields are shown in linear scale, with color code ranging from -150 (black) to +150~km~s$^{-1}$ (white) for the velocity maps, and from 0 (black) to 200~km~s$^{-1}$ (white) for the velocity dispersion. The contour intervals are 35 km~s$^{-1}$ in the velocity and dispersion fields. The system is viewed under the projection that maximizes its apparent flattening parameter $E$ at each timestep: other projections make the system appear rounder, with generally lower rotational velocites.}
\label{snapshots}
\end{figure*}

We show in Fig.~\ref{snapshots} the morphology and kinematics of the {\tt 1x10:1} to {\tt 4x10:1} merger remnants. After one merger of this mass ratio, the system resembles an S0 galaxy. It is made up of a massive disk, thickened by the interaction, and has large velocity dispersions but still with $V/\sigma>1$. After three or four 10:1 mergers, the remnant does not resemble a disk galaxy any longer: it is a spheroidal galaxy, with maximal flattening E5--E6 (but less in most projections). The isophotes have only a low residual diskiness, and the Sersic index $n \simeq 3$ shows that the radial luminosity profile is now closer to an ``$R^{1/4}$'' profile (de Vaucouleurs 1964) than exponential. The morphology of this system is hence typical of an elliptical galaxy. At the same time, the kinematics is dominated by velocity dispersions with $V/\sigma = 0.6$ after four mergers, similar to what is found in remnants of major 3:1--2:1 mergers (paper~I and references in the introduction). 

The remnant of the succession of four 10:1 mergers has thus both the morphology and the kinematics of an elliptical-like galaxy, while each of these mergers would individually result in an early-type disk galaxy. After the {\tt 4x10:1} merger sequence, the system has the properties of a disky elliptical galaxy where a significant residual rotation is still observed. With an increased number of minor mergers it can become a boxier, slower-rotating elliptical galaxy (see. Fig~\ref{seq_a4}). Also note in Fig.~1 that the residual rotation axis also tends to become misaligned to the apparent flattening direction with increasing number of mergers. 

\subsection{Morphology and kinematics of the multiple merger remnants}

\begin{figure}
\centering
\resizebox{8cm}{!}{\includegraphics{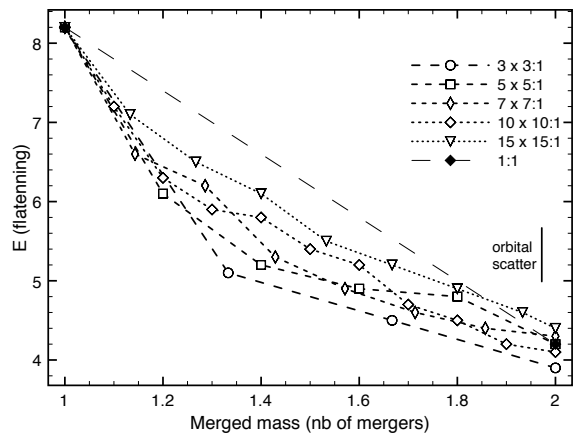}}
\caption{Evolution of the apparent flattening of the relaxed merger remnants seen edge-on, along sequences of mergers with various mass ratios. All systems evolve from the initial thin disk to spheroidal galaxies. The flattening appears to depend on the merged mass, more than the mass ratio of the mergers: for instance 6 mergers of mass ratio 10:1 or 3 mergers of mass ratio 5:1 (that both correspond to a merged mass of 1.6) both lead to the formation of an E5 system. Only the 15:1 systematically forms flatter systems, but the difference remains small and several mergers of this type still form spheroidal galaxies. The values for 1:1 and 3:1 mass ratios are averaged between prograde and retrograde orbits as detailed in the text. The ``orbital scatter'' shows the ($1 \sigma$) dispersion of the result for the {\tt 10x10:1} sequence when orbital parameters are varied (as detailed in Sect.~\ref{411}).}
\label{seq_E}
\end{figure}

\begin{figure}
\centering
\resizebox{8cm}{!}{\includegraphics{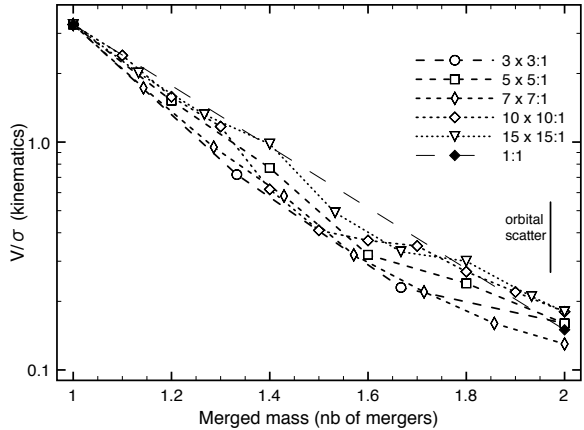}}
\caption{Evolution of the $V/\sigma$ parameter along sequences of mergers of various mass ratios. All systems evolve from the initial rotating disk to remnants supported by increasing velocity dispersion. The kinematics of these multiple merger remnants appears to depend on the merged mass more than the mass ratio of the mergers. The velocity dispersion becomes larger than the rotation velocity when the merged-mass is larger than $\sim$1.3, after for instance one single 3:1 merger or three mergers of mass ratio 10:1. Very slow-rotating systems are obtained once the mass of the system has doubled via the succession of mergers, as after one single equal-mass merger. The ``orbital scatter'' is as in Fig.~\ref{seq_E} (see also Sect.~\ref{411}).}
\label{seq_VS}
\end{figure}

\begin{figure}
\centering
\resizebox{8cm}{!}{\includegraphics{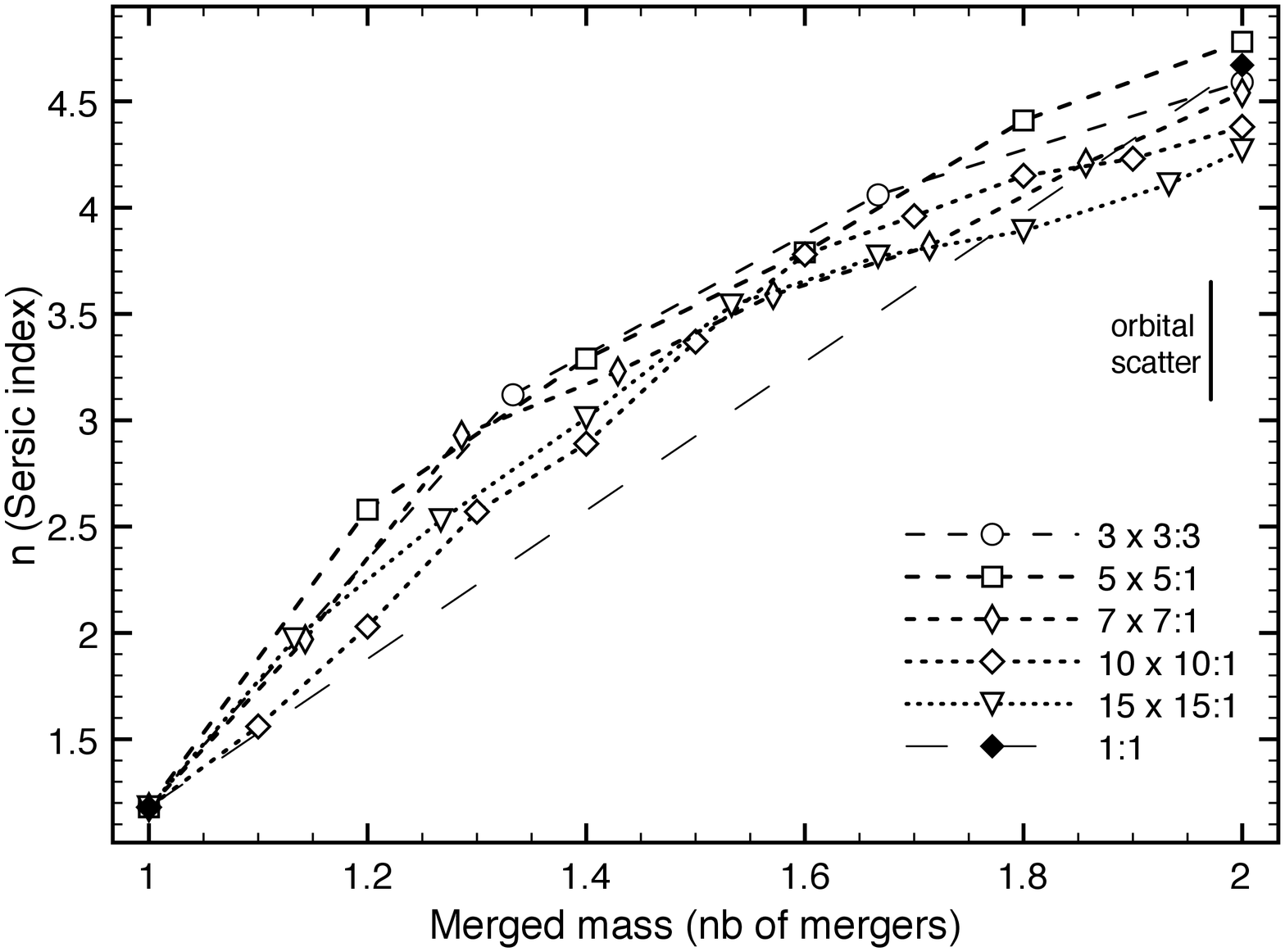}}
\caption{Evolution of the Sersic index $n$ in the $[0.55R_{25};R_{25}]$ radial range. Starting from $n \sim 1$ for the initial disk, its value increasing while the luminosity profile becomes more typical of an elliptical galaxy. The ``orbital scatter'' is as in Fig.~\ref{seq_E} (see also Sect.~\ref{411}).}
\label{seq_n}
\end{figure}

The sequences of mergers with various mass ratios have been systematically analyzed. The fundamental parameters $E$, $n$, $V/\sigma$, as defined in Sect.~2, are plotted in Figs.~\ref{seq_E}, \ref{seq_VS} and \ref{seq_n} for each sequence. The results are given as a function of the number of mergers converted into the {\emph{``merged mass''}}: for instance, when the initial galaxy has merged with 3 companions, each of them having a 5:1 mass ratio, this so-called merged mass is 1.6 (1 for the main initial galaxy and 0.2 for each companion). Using this definition allows us to directly compare sequences with various mass ratios. We recall that each multiple merger remnant is analyzed after 1.5~Gyr of evolution to make sure that the system is reasonably relaxed. We do not analyze the poorly relaxed systems between the occurrences of two successive mergers or the on-going merger properties in this paper.

The first noticeable result is that once the merged mass is 2.0 (i.e., the total mass of the companions equal the mass of the initial galaxy, as in a 1:1 major merger), the system has the properties of a slow-rotating elliptical-like galaxy, be the companions massive (like 1:1 and 3:1) or much more minor (like 15:1). Indeed, all the final systems after these sequences of mergers have:
\begin{itemize}
\item a {\emph{spheroidal shape}} with an axis ratio larger than 0.5 (class E4 or at most E5) when observed ``edge-on''. The shape of the systems is even rounder with other lines-of-sight. This is measured in the outer parts of the galaxies (up to $R_{25}$), so that the systems cannot simply consist of spheroidal bulges surrounded by thin disks, but are really spheroid-dominated galaxies -- if a thin outer disk is present, it lies beyond the 25th isophote and/or does not dominate the mass distribution.
\item a {\emph{Sersic index of 4.5}} on average, thus being much closer to the empirical ``$R^{1/4}$'' law (de Vaucouleurs 1964) than to the exponential profile of a disk galaxy. Here again the measurement is not restricted to the central regions but relates to the radial profile up to the 25th isophote.
\item a kinematical parameter {\emph{$V/\sigma$ smaller than 0.3}} (on average 0.2), and are hence not rotationally-supported. Random motion dominates in these multiple merger remnants as in real elliptical galaxies, and are even larger than in the remnant of one single 3:1 merger (which is usually already considered to be an elliptical-like system).
\end{itemize}

There are differences from one case to the other, but they are not larger than the variations from one line-of-sight to another one for each given system, and much below the differences between standard elliptical and spiral galaxies. The main observable properties, in projection, of these remnants from multiple mergers are typical of elliptical galaxies, even if each merger is really minor (like 10:1 or 15:1) and would by itself alone have kept the system in the form of a (disturbed) spiral galaxy.

A merged mass of 2.0 is not required to form an elliptical-like galaxy. This is known to be the case for the ``major'' unequal-mass 2:1 and 3:1 binary mergers. A set of criteria to consider that a merger remnant is an elliptical galaxy can be:
\begin{itemize} 
\item a flattening of E6 or less when observed edge-on, i.e. thicker than any disk galaxy.
\item a kinetic energy mainly in the form of random motion, i.e. $V/\sigma<1$.
\item a Sersic index larger than 3, i.e. the luminosity profile is closer to elliptical-like than to an exponential disk -- an this cannot relate only to a central bulge since the inner regions are not included in our measurement. 
\end{itemize}
Thus, elliptical-like systems are obtained as soon as the merged mass is 1.3--1.4, i.e. the total mass of the companions is at least 30\% of that of the initial galaxy. This is in agreement with binary mergers up to 3:4 or 4:1 already producing elliptical galaxies (Bournaud et al. 2005b -- potentially with faint outer disks: Naab \& Trujillo 2006), but this also indicates than two mergers with mass ratio 5:1 or three mergers with mass ratio 10:1 are enough to transform a spiral galaxy into an elliptical galaxy. These systems then resemble the low-mass elliptical galaxies that have a higher degree of residual rotation than the giant elliptical galaxies (e.g., Naab \& Burkert 2003). An increasing number of mergers finally keeps on evolving them into more concentrated and rounder\footnote{these systems are rounder regarding their global flattening $E$, but this does not imply that the diski-/boxiness is washed out (see Sect.~3.3).} systems, together with a still decreasing degree of rotation.

\smallskip

Thus, multiple minor mergers can form elliptical galaxies that look like both the real observed ellipticals and the major merger remnants, from their shape (flattening), radial profile (Sersic index) and kinematics ($V/\sigma$). These elliptical-like galaxies are never flatter than E7, like the real ellipticals and the major merger remnants. This is because the increase of the equatorial velocity dispersion is not achieved without a large increase of the vertical dispersions, too, which makes the final system much thicker than its progenitors. 
 
With the parameters studied so far, no major systematic difference exists compared to major mergers, provided that the total merged mass is the same. This can be interpreted as follows: the energy dissipated by dynamical friction is to first order equal to the relative kinetic energy of the merging galaxies (i.e. their mass and relative velocities). The merging of an equal-mass pair thus releases the same energy through dynamical friction as ten subsequent 10:1 mergers. The kinetic energy dissipated by dynamical friction heats the stellar systems. The same amount of heating is thus obtained in the two cases (binary 1:1 vs. sequence of ten 10:1 merger), which explains the similar $V/\sigma$ final values. Kinematical heating in galaxy mergers is generally close to isotropic, thus the different cases also lead to the same amount of vertical heating of the initial disk, which explains the similar flattening of the merger remnants of various origins. Finally, the rather similar Sersic indexes suggest that the final degree of relaxation is similar, even though it is reached through gradual stages. A detailed study of the dynamics and stellar orbits during on-going mergers will however be needed to fully understand the similarities and differences between the violent relaxation of major mergers and the step-by-step relaxation through multiple minor mergers.

\subsection{Boxiness of elliptical galaxies}

Elliptical galaxies are often classified according to various parameters. A fundamental one, that relates to their formation history, is the diskiness/boxiness parameter $a_4$, which quantifies the 4th Fourier component\footnote{We define $a_4$ as the coefficient of the $\cos \left( 4\theta \right)$ component in the azimuthal decomposition of the isophotes.} in the azimuthal decomposition of their projected isophotes \citep[e.g.,][]{BM87}. Two classes of elliptical galaxies are generally distinguished:
\begin{itemize}
\item the ``disky'' elliptical galaxies ($a_4>0$) are not disk galaxies but have an isophotal shape recalling that of disks, although at a much lower degree. These are generally low mass elliptical galaxies, with a significant amount of rotation ($V/\sigma$ up to 0.5--1). In binary merger scenarios, these are mainly produced by 3:1--4:1 mergers.
\item the ``boxy'' elliptical galaxies ($a_4<0$), whose isophotes have a distorsion that recalls a box (or a rectangle) compared to perfect ellipses . These are generally higher mass elliptical galaxies, with less rotation ($V/\sigma$ generally limited to 0.1 or 0.2). In binary merger scenarios, they result mainly from equal-mass 1:1 mergers and sometimes from 2:1 mergers. But these equal-mass binary mergers fail to reproduce the most boxy elliptical galaxies, in particular the giant elliptical galaxies \citep[e.g.,][]{naabburkert03,naabostriker07}.
\end{itemize}

\begin{figure}
\centering
\resizebox{8cm}{!}{\includegraphics{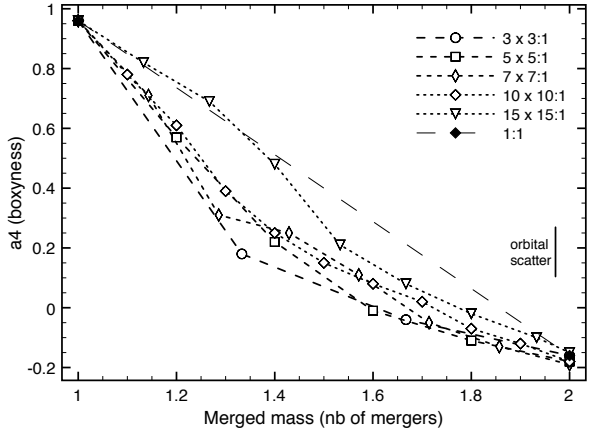}}
\caption{Evolution of the $a_4$ parameter versus the merged mass along several merger sequences. The elliptical-like galaxies formed with merged masses up to $\sim 1.6$ have a disky isophotal shape (on average), but the high diskiness of the initial galaxy is washed out by the multiple minor mergers. When the total merged mass increases, the remnants tend to show boxy isophotes (on average). The ``orbital scatter'' is as in Fig.~\ref{seq_E} (see also Sect.~\ref{411}).}
\label{seq_a4}
\end{figure}

We have measured the $a_4$ parameter for the multiple merger remnants from several projected images (same as $V/\sigma$). The average results for the various simulated mass ratios are shown in Fig.~\ref{seq_a4}. Once again, the result depends mainly on the total merged mass with little dependence on how it is merged. For instance, two or three successive mergers of mass ratio 7:1 produce the same final low diskiness as one single 3:1 merger, while 10 minor 10:1 mergers result in a boxiness comparable, on average, to an equal-mass 1:1 merger. Note that, just like a {\tt 1:1} remnant, a {\tt 10x10:1} remnant can appear slightly disky depending on the viewing direction (see the 1-$\sigma$ scatter in Fig.~\ref{seq_a4}) but will more frequently appear boxy, with on average $a_4 \sim$-0.1~--~-0.2. Nevertheless, a trend can be noted for lower mass ratios, in particular 10:1 and 15:1, to converge less rapidly towards negative $a_4$ parameters. This indicates that the less violent relaxation in minor mergers is less efficient in destroying the disky underlying orbital structure. The difference remains small after several 10--15:1 mergers, but this suggests that very minor mergers at higher mass ratios might be less efficient in forming boxy elliptical galaxies. This trend could also be increased by the higher gas mass fraction in lower mass galaxies, because this gas tends to reform disky structures in elliptical galaxies, which can reduce the boxiness.

Our result extends the trend observed in 1:1 vs 3:1 binary mergers, namely an increase in the boxiness with the merged mass. This trend still holds when several mergers are repeated in time. Thus, boxy elliptical galaxies can be formed either in the merging of pairs of nearly-equal-mass galaxies, but also by the merging of several lower mass galaxies. \citet{limaneto} had found that repeated mergers tend to wash out any particular shape (boxy or disky) of elliptical galaxies. The apparent contradiction can relate to the much lower resolution available to these authors, but also that they studied only collisionless and simultaneous major mergers, which is a largely different process.

We actually find here that an increasing number of subsequent mergers does increase the boxiness. In the same vein, (Naab, Khochfar \& Burkert 2006b) find that an elliptical galaxy that undergoes another merger (with another elliptical galaxy in their simulations) can have a slight increase in its boxiness, too. Binary mergers of disk galaxies cannot account for the most massive elliptical galaxies being boxy: even collisionless, perfectly equal-mass 1:1 mergers generally do not result in very boxy systems. The formation of such giant boxy ellipticals can then more likely result of from several re-mergers, the first ones forming an elliptical galaxy and the last ones making it more and more boxy. This mechanism is at work in the simulations by \citet{naab06early}, and our results (Fig.~\ref{seq_a4}) further indicate that the sequential mergers do not need to be equal-mass nor even major : several 3:1 mergers can form a boxy elliptical galaxy, and a sufficient number of 7:1 mergers can as well. An alternative scenario to form massive boxy elliptical galaxies might be the simultaneous merger of several galaxies in a dense proto-cluster environment, as in the simulations by \citet{WeilH94,WeilH96}: the morphological properties of massive elliptical galaxies formed this way remain to be studied in detail. We at least show that repeated mergers can explain the boxiness of large ellipticals, which binary major mergers of disks alone fail to explain. It is possible that the giant ellipticals first formed from major mergers at high redshift, then continued to grow with repeated minor mergers with lower-mass galaxies, hence gradually increasing their boxiness.

\subsection{Anisotropy of stellar orbits and the dark matter content of elliptical galaxies}

Up to now, we have analyzed only the projected properties of the multiple merger remnants. These can be directly measured in observed galaxies. The three-dimensional orbital structure of elliptical galaxies cannot be directly observed, but there is some evidence that the rotation of elliptical galaxies is generally not large enough to support their flattening \citep{binney82}, so that their stellar velocity dispersion has to be anisotropic. Yet, the actual anisotropy of the orbits cannot be accurately inferred from observable parameters alone. And indeed, it is only recently that numerical simulations of major mergers revealed the large radial anisotropy of the stellar orbits in elliptical galaxies, in particular in the outer regions. This anisotropy results in a decrease of the observed velocities (along the line-of-sight) in the outer regions, so that the actual mass content of elliptical galaxies had been underestimated -- in particular their dark matter content \citep{dekel2005}.

\begin{figure}
\centering
\resizebox{8cm}{!}{\includegraphics{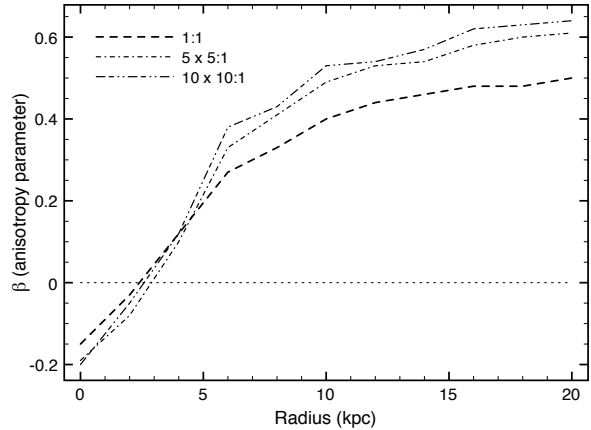}}
\caption{Anisotropy parameter $\beta$ for an equal-mass major merger remnant (1:1) compared to remnants of multiple minor mergers ({\tt 5x5:1} and {\tt 10x10:1}). The radial variations of the anisotropy are qualitatively similar in these systems, but the remnants of multiple minor mergers have a higher radial anisotropy in their outer regions.}
\label{aniso}
\end{figure}

In the multiple merger remnants, we measured the anisotropy parameter $$\beta = 1 - \left( \frac{\sigma_{\theta}^2}{\sigma_{r}^2} \right)$$ where $\sigma_{\theta}$ and $\sigma_{r}$ are respectively the tangential and radial velocity dispersions with respect to the galaxy mass center. The result is shown, on average over all directions but as a function of radius, in Fig.~\ref{aniso}. The elliptical-like galaxies formed in multiple minor mergers tend to have a larger radial anisotropy than those formed in one single merger bringing the same total mass.

A possible explanation of the higher radial anisotropy found in multiple minor merger remnants can be that the stars and gas clouds in smaller companions have a lower angular momentum than in a massive one. Then, the resulting remnants have a lower angular momentum per star, which results in more eccentric orbits, i.e. higher radial velocity at a given radius. As a result, higher radial anisotropies would tend to increase the dynamical mass of elliptical galaxies \citep{dekel2005} if some of them have been formed by multiple minor mergers, but the order of difference expected from Fig.~\ref{aniso} would not exceed a few tens of percent. However, a study including various orbits and initial morphologies is required to confirm whether the anisotropy is systematically higher in multiple minor merger remnants. Indeed, \citet{athan05} shows that many parameters can significantly influence the orbital structure of the merger remnant, including the time interval between subsequent collisions in group mergers. Moreover, the radial anisotropy is unlikely to keep increasing with higher numbers of minor mergers, given that \citet{athan05} finds lower anisotropy in remnants whose progenitors are already elliptical.

\section{Discussion}

\subsection{Formation of elliptical galaxies without major mergers}\label{41}

\subsubsection{Remnants from multiple minor mergers and orbital parameters}\label{411}

Single binary mergers with mass ratios of 5:1 and larger do not form remnants with elliptical-like properties, but disk galaxies: these are ``minor'' mergers, or the ``intermediate'' mergers forming S0-like galaxies (Bournaud et al. 2005b). But when several mergers of this type occur, the system shows increasing velocity dispersion, together with a rounder and more concentrated morphology (lower $E$ and higher $n$). Hence, both the morphological and kinematical properties of these multiple minor merger remnants are similar to those of observed elliptical galaxies. An elliptical-like galaxy is formed when the total mass of the companions is 30--40\% of the mass of the main initial galaxy (merged mass = 1.3--1.4). This has been established for 5:1 to 15:1 companions, but the simulation with 50:1 companions presented below indicates that this result extends to smaller companions, too. Multiple minor mergers are thus a new pathway for the formation of elliptical galaxies without major mergers. 

The sequences of mergers in our simulations described so far alternate prograde and retrograde companions. The impact parameter, velocity, and inclination were not fixed at particularly low or high values, and are thus expected to give representative results. Furthermore, we know from paper~I and other studies of binary mergers that the global properties of a merger remnant are more influenced by the mass ratio of the merging galaxies than the other parameters. For instance, the properties of a binary 3:1 merger remnant vary with the orbit, but the variations are generally smaller than the typical differences with binary 1:1 or 5:1 remnants. Yet, the influence of orbital parameters may a priori be more important for multiple mergers, because the total angular momentum provided to the system can change, influencing the residual rotation of the merger remnant. 

We have then repeated our {\tt 10x10:1} simulation with all companions on prograde orbits, all companions on retrograde orbits, all companions at inclination $\theta = 0$ (alternating prograde and retrograde orbits as in the fiducial run), and all companions at $\theta = 65$ degrees. The typical variation of the various properties $E$, $V/\sigma$, $n$ and $a_4$ corresponding to these changes in orbits is shown in Figs.~\ref{seq_E} to \ref{seq_a4}. We then notice that: {\emph{(i)}} for a given merged mass (for instance {\tt 10x10:1} vs {\tt 3x3:1}) the differences between the mass ratios are below this orbital scatter, but {\emph{(ii)}} the evolution of the morphological parameters along the merger sequences is larger than the scatter related to orbital parameters, hence it is a significant general result. The fact that a {\tt 10x10:1} sequence produces slow-rotating and (slightly) boxy ellipticals, on average, while a {\tt 6x10:1} sequence results in a disky rotating elliptical is a robust result that is not much affected by the variations of orbital parameters. Only the case where all the mergers occur on prograde orbits gives a somewhat higher $V/\sigma$, but this is a rather unlikely case. In major mergers too, some orbits can preserve massive disk-like structures \citep[for instance in the case of NGC~4550,][]{pfenniger4550}. Similarly, a succession of ten minor mergers with all companions coplanar and corotating to the main galaxy would likely form an S0 or disky E rather than a boxy E, but such unlikely cases can only explain specific situations without being representative of the majority of real mergers. 

\begin{figure}
\centering
\resizebox{8cm}{!}{\includegraphics{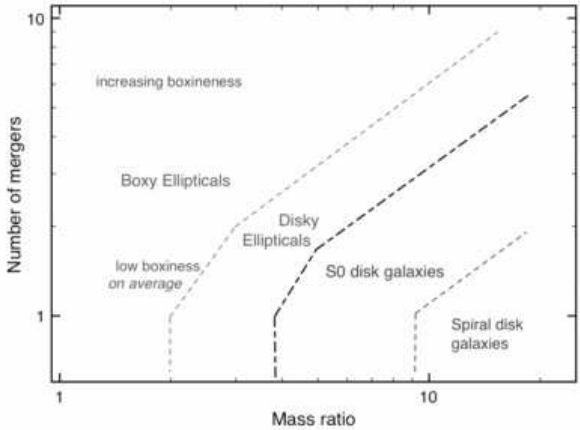}}
\caption{Schematic description of the nature of the binary and multiple merger remnants, as a function of the number of mergers and the mass ratio. Orbital parameters have not been varied in this paper, except for 10:1 mergers, but we know from paper~I that the properties of a merger remnant depend on the mass ratio more than the orbital parameters, so this scheme should give a representative estimate of the number of mergers required to transform a disk galaxy (Sp or S0) into an elliptical-like system. The ellipticals formed for instance by {\tt 1:1} mergers or {\tt 5~x~7:1} sequences are generally boxy {\emph{on average}} (they appear disky as well under some projections), and their boxiness is moderate. An increasing degree of boxiness can be obtained with an increasing number of mergers. These results were established for a gas mass fraction of 15\% within stellar disks but to first order can be applied to all systems that are not initially strongly gas-dominated (see text).}
\label{params}
\end{figure}

In Fig.~\ref{params}, we briefly summarize the properties of the binary and multiple merger remnants, as a function of the mass ratio and number of mergers, according to the simulations of this paper and that of paper~I. Real situations can obviously be more complex, but from what we said above, we can expect for instance that a {\tt 5:1-10:1-10:1} sequence will form an elliptical-like galaxy whose average properties are similar to the {\tt 4x10:1} and {\tt2x5:1} remnants.

\subsubsection{Higher mass ratios: can very small companions still form ellipticals?}\label{412}

Here we present a simulation where the same initial spiral galaxy as before merges with 50 companions each of 1/50th of its initial mass. The parameters are the same as for the other simulations, except that:\\
-- the mergers cannot be fully subsequent; they are uniformly distributed over a 8~Gyr period.\\
-- the dwarf companions initially contain 30\% of their visible mass in gas.

\begin{figure*}
\centering \sidecaption
\resizebox{12cm}{!}{\includegraphics{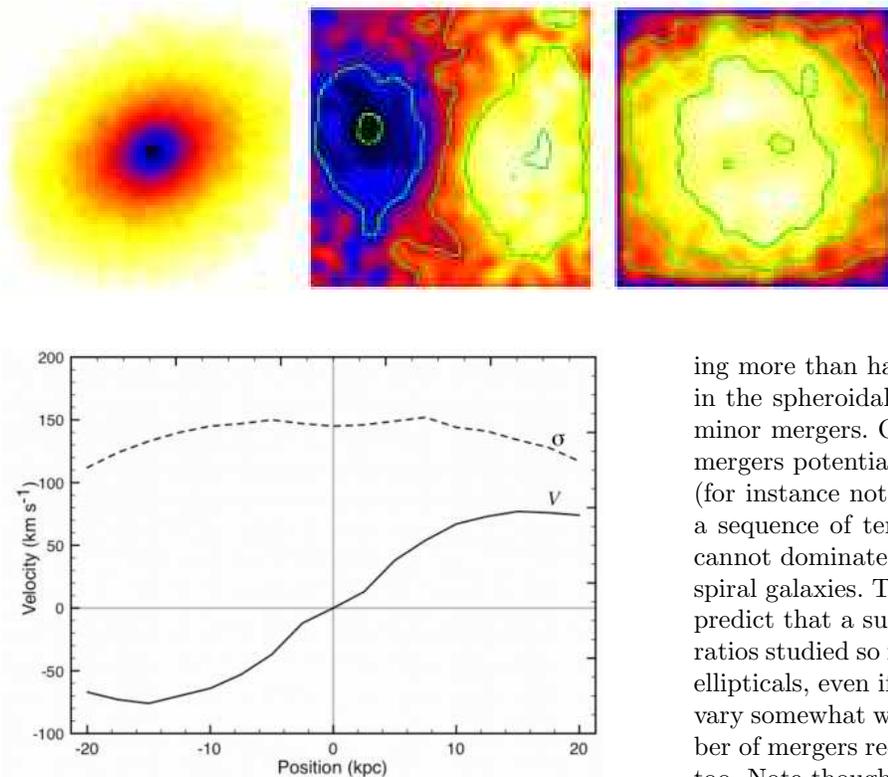}}
\caption{Projected mass density, velocity and dispersion fields for a relaxed {\tt 50x50:1} merger remnant, viewed along the projection resulting in the largest apparent flattening ($E=6.1$). The contours on the velocity field are 35 and 70 km~s$^{-1}$, on the velocity dispersion field they are 100, 120 and 140~km~s$^{-1}$. Note the misalignment between the residual rotation and apparent flattening axis.}
\label{sim_50}
\end{figure*}

\begin{figure}
\resizebox{8cm}{!}{\includegraphics{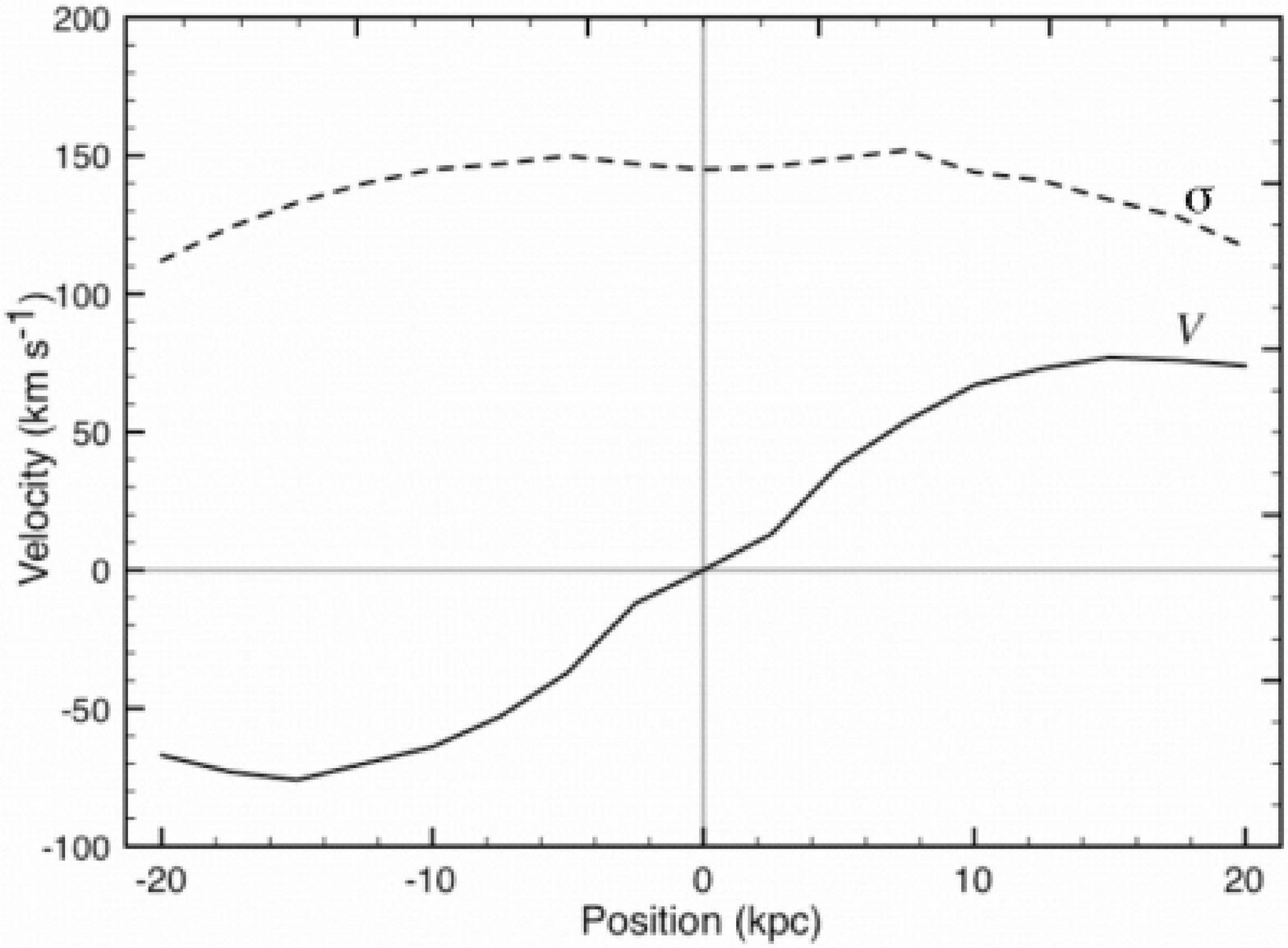}}
\caption{Average velocity and velocity dispersion for the {\tt50x50:1} merger remnant shown in Fig.~\ref{sim_50}, taken along a 1 kpc-width slit aligned with the apparent major axis. The system is supported by large velocity dispersions similar to those observed in elliptical galaxies.}
\label{sim_50_rc}
\end{figure}

The relaxed remnant of this {\tt 50x50:1} multiple minor merger sequence is displayed in Fig.~\ref{sim_50}, viewed along the line-of-sight giving the largest apparent flattening (which is $E=6.1$). This system has the shape of an E6 elliptical galaxy (but would more likely be observed as an E5 to E0 one under other projections), and is dominated by velocity dispersions. We show the mean velocity and velocity dispersion profiles that would be observed along the major axis of this system in Fig.~\ref{sim_50_rc}, which confirms that it has elliptical-like large-scale kinematical properties with $V/\sigma \sim 0.5$ even in the outer regions.

The remnants of 5:1 to 15:1 multiple mergers did not show a large variation of their properties with the mass ratio at given total merged mass. This already suggests that the result was the same for a higher number of smaller companions. The result of the {\tt 50x50:1} confirms that the formation of elliptical-like galaxies can still be achieved with high mass ratios. In detail the {\tt 50x50:1} merger remnant is slightly flatter and has a somewhat higher residual rotation than the other cases; the discrepancy is not major and likely results from the higher gas fraction in the 50:1 companions. Thus, a large number of mergers with very high mass ratios still form elliptical-like galaxies, at least when the size/concentration of galaxies is scaled as in our model (which assumes $M \propto R^2$ and a constant central surface density). Resolution tests should be performed in the future to further explore the case of such very small companions. A few $10^5$ particles per galaxy is usually believed to provide a viable large-scale description for major mergers, but in the 50:1 cases the small companion galaxy contains only $2\times 10^4$ stellar particles. High mass ratios are difficult to model for this reason, and a lack of resolution could for instance smooth their gravitational potential and underestimate their effects.

\subsubsection{Mergers of gas-rich and high-redshift galaxies}\label{413}

The initial gas mass fraction in the initial spiral galaxy and in the 1:1 to 15:1 companions is 15\% in our models: the mergers are not ``dry'', but higher gas fractions could still be encountered in real mergers at high redshift and/or in low-mass companions. A high fraction of gas in mergers tends to preserve the diskiness and somewhat higher residual rotation (Naab, Jesseit \& Burkert 2006a; Jesseit et~al. 2007) because the gas can be stripped before the galaxy collision, making it less violent, and also because gas can fall back into a disk where it will form new stars after the merger. However, major mergers rarely preserve disk galaxies similar to real spirals. Indeed, even cases starting with pure gas disks in a coplanar co-rotating situation (the best case to preserve most of the angular momentum) end up with about half the mass in a spheroidal component \citep[see e.g.,][]{SH05}. More realistic orbits will have more mass in the spheroid, and more realistic initial gas fractions will reduce the subsequent disk re-formation so that even more mass will end up in a spheroid, making the galaxy elliptical-like (there can be an outer disk component, formed for instance by gas falling-back, but not dominating the mass). So gas-rich major mergers are still expected to form elliptical galaxies, even if their detailed properties show differences. Only extreme situations with very high initial gas fractions can preserve an (early-type) disk (see also \cite{robertson06}), which can be representative only of primordial galaxies at very high redshift, but not of most galaxies at $z\sim1$. Our simulations, even with 30\% gas for the 50:1 case, can have a low-mass disk component but not a final mass distribution dominated by an exponential disk.

As for multiple minor mergers, except at very high redshift $z>>1$, gas mass fractions are not expected to be larger than 50\%, except perhaps in some exceptional cases or in LSB galaxies. Then the pre-existing stars, representing more than half of the mass, will to first order end-up in the spheroidal component after a sufficient number of minor mergers. Only the stars born during and after the mergers potentially end up in a disk, but not all of them (for instance not those formed during the first merger in a sequence of ten mergers), so that the disk component cannot dominate the mass distribution -- which it does in spiral galaxies. Thus, at redshift $z\sim1$ and below, we can predict that a succession of minor mergers with the mass ratios studied so far will still transform spiral galaxies into ellipticals, even if the detailed properties of the latter can vary somewhat with the gas fractions, and the exact number of mergers required to make an elliptical could change too. Note though that high-redshift disks frequently have peculiar morphology with sometimes much mass in giant clumps (Conselice 2003; Elmegreen, Elmegreen \& Hirst 2004; Conselice et~al. 2004; Elmegreen 2007; Bournaud, Elmegreen \& Elmegreen 2007), and the potential influence of such properties on the structure of high-redshift major and minor merger remnants remains to be explored.

\subsubsection{Comparison with major mergers and star formation history}\label{414}

The global properties of the multiple minor merger remnants resemble those of binary major mergers. It is thus difficult to infer from the projected properties of an elliptical galaxy its past formation mechanism. The differences between multiple minor and major mergers are not larger than the scatter related to orbits and other parameters. The orbital structure can be different, but it is unclear whether there is a systematic difference, because many parameters that may also influence the anisotropy of the merger remnants  have not been varied in this paper.

Differences may more likely be found using more detailed parameters, like the $h_3$ of the projected velocity distribution (Gonz{\'a}lez-Garc{\'{\i}}a, Balcells \& Olshevsky 2006), the $\lambda_R$ parameter defined by \citet{emsellem07} which couples the resolved kinematics to the density profile and revealed different families of early-type galaxies, or correlations of several such parameters. Such detailed comparisons with advanced parameters are beyond the scope of the present paper which studies whether merger remnants are elliptical-like or not, but will be important for future studies.

\begin{figure}
\resizebox{8cm}{!}{\includegraphics{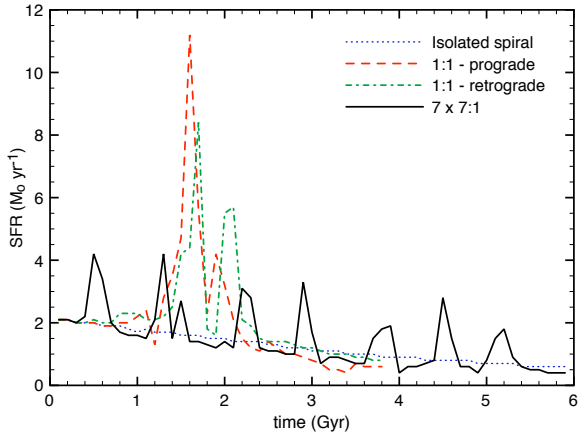}}
\caption{Star formation history in binary major mergers (1:1, for the prograde and retrograde orbits) and for a succession of 7:1 minor mergers.}
\label{sfr}
\end{figure}

Another major difference lies in the star formation history of these systems. Indeed, minor mergers trigger bursts of star formation \citep[e.g.,][]{coxphd, cox07} that are comparatively less intense than in major mergers \citep[see][regarding the intensity of starbursts in major mergers]{dimatteo07}. As a result, the star formation history of multiple minor merger remnants largely differs from that of a binary major merger, as illustrated in one case in Fig.~\ref{sfr}. This difference may however become smaller if the successive minor mergers occurred with smaller time intervals or simultaneously.

\subsubsection{Scaling relations in elliptical galaxies}

Beyond their fundamental properties studied above (mass profile, kinematics, isophotal shape), the real elliptical galaxies are also observed to be described by relations between these fundamental parameters. An important scaling relation is the so-called fundamental plane \citep{faber87}. Most scaling relations are defined over the complete mass (or luminosity) range of elliptical galaxies, while our simulations explore only a restricted mass range (a factor of two between the initial and final masses), making it irrelevant to test the viability of multiple minor mergers by a comparison with these relations. This is also the case for major merger simulations, and simulations covering the full mass range of ellipticals in a cosmological context would be required.

\begin{figure}
\resizebox{8cm}{!}{\includegraphics{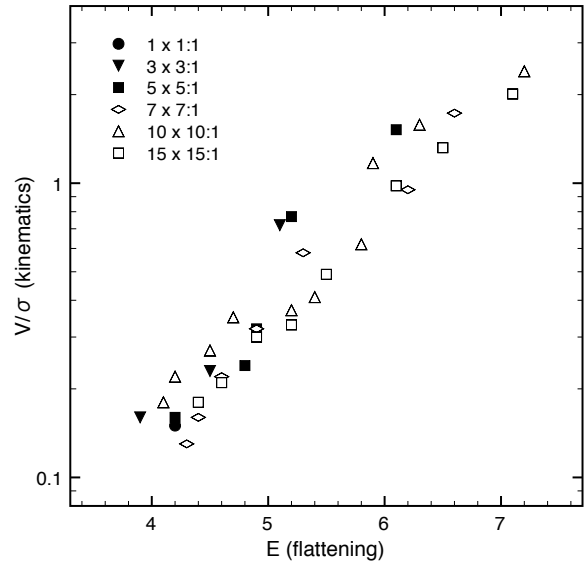}}
\caption{Relation between the velocity dispersion (measured via the $V/\sigma$ parameter) and ellipticity (flattening $E$) along the multiple minor merger sequences with various mass ratios, and the reference major mergers. Evolution with increasing number of mergers is from top-right to bottom-left.}
\label{b_rel}
\end{figure}

An observed fundamental relation of ellipticals that can be tested in our simulations is the ellipticity -- velocity dispersion ($E , V/\sigma$) relation \citep{binney82}, which does not explicitly span a large mass range. In our simulations of multiple minor mergers, we find in Fig.~\ref{b_rel} that a rather tight relation exists between $E$ and $V/\sigma$ for the elliptical remnants of multiple minor mergers -- and even for those that we classify as disks (spirals or S0s) after for instance one single 7:1 merger or two 15:1 mergers. The multiple merger mechanism then appears to form elliptical galaxies that are viable with regard to this observed relation. The fact that all fundamental parameters have a similar evolution along all mergers sequences, whatever the mass ratio is (Figs.~2 to 5), suggests that other parametric relations exist too. Note however that we cannot check in our simulations whether higher- or lower-mass ellipticals would still lie on the relation shown in Fig.~\ref{b_rel} and not offset from it.

The mass of spheroids (bulges and ellipticals), often traced by central velocity dispersion, is known to correlate with the mass of central black holes \citep[e.g.,][]{KR95}, probably mostly grown during rapid accretion QSO phases \citep{YuT02}. Major mergers can directly fuel a central black hole by rapid gas accretion \citep{TDiM07}, while minor mergers would rather lead to more moderate inflows (see the star formation history in major vs multiple minor mergers in Fig.~\ref{sfr}). Still, minor mergers can increase the mass of a central black hole by the merging of several smaller black holes - those initially at the center of the primary target galaxy and each companion: then the relation between the central black hole mass and the spheroid mass could a priori be preserved in the elliptical galaxies formed by multiple minor mergers.

\subsection{Constraints on galaxy growth and evolution}

Repeated minor mergers, even with rather high mass ratios, are thus a mechanism that can form spheroids having the properties of the real observed elliptical galaxies. Multiple minor mergers is then a possible mechanism to form elliptical galaxies without major mergers, at least within the {\emph{theoretical}} frame where binary and repeated mergers plus internal evolution are the only phenomenon driving the evolution of galaxies. The {\emph{real}} evolution of galaxies is more complex, because galaxy-galaxy mergers and internal evolution are not the only driving mechanisms, and we now discuss how repeated minor mergers can be placed in a more complete description of galaxy evolution.

\subsubsection{The survival and growth of spiral disks}

Disk galaxies are observed to increase their size and mass between $z=1$ and $z=0$. At redshift 1, stellar disks had on average smaller truncation radii \citep{trujillopohlen05} and smaller exponential scalelengths \citep[e.g.,][]{elm07}. Their mass must then have increased in significant proportions over the last $\sim 8$~Gyr. If this was to be achieved mainly through minor mergers, most spiral galaxies would have been transformed into S0 (mass increased by $\sim$20--30\% through mergers) and ellipticals (mass increased by 30--40\% and more). It is then required that another process participates in the growth of spiral galaxies, likely the accretion of external gas. While minor mergers progressively destroy disks, accretion of gas can refuel massive thin disks to maintain high disk-to-bulge mass ratios. It can be the accretion of diffuse gas falling into the potential well of galaxy halos, or the accretion of gas-dominated clumps with low visible and dark masses, that are disrupted in the early stages of the interaction. But usual dwarf galaxies of M$_{\mathrm{vis}} \sim 10^9$~M$_{\sun}$ and above plus their massive dark matter halo cannot achieve this, as shown for instance by our simulation with 50:1 mergers. 

Cold gas can be accreted directly from cosmological filaments for galaxies whose total mass (including the dark halo) does not exceed $\sim 10^{12}$~M$_{\sun}$ \citep{birnboimdekel03, dekelbirnboim06} and even for more massive galaxies but less regularly (Birnboim, Dekel \& Neistein 2007); and it can explain the morphological properties of disk galaxies in our Local Universe \citep{BC02, bournaud05m1}. Moreover, the growth of disks is observed to be inside-out \citep[e.g.,][]{trujillopohlen05}, which cannot be easily explained by a (minor) merger-driven growth, but can result from the accretion of gas if the lower angular momentum material is accreted first, and higher angular momentum gas accreted at later stages. Gas falling-back from large radii after the mergers (in particular gas from tidal tails) appears not to be massive enough to refuel a massive disk, even in our 50:1 cases with gas-rich companions. A low-mass disk component can be fueled this way, but a disk dominating the total mass distribution cannot be maintained without the additional accretion of external gas.

That spiral disks have been able to grow from $z=1$ to $z=0$ thus implies that some phenomenon favor the persistence of their disks, like the accretion of cold gas. Then, the process of progressive disk destruction and transformation into ellipticals by repeated minor mergers could be compensated for by other processes, like cold gas accretion, which would increase or maintain massive disk components around/within the heated stellar spheroid. But in our simulations the galaxy evolution is driven only by minor mergers, and this leads to a progressive transformation of disk galaxies into elliptical-like spheroids.

\subsubsection{The number of elliptical galaxies and their formation}

The standard ``major merger'' scenario for the formation of elliptical galaxies is usually assumed to produce a fraction of elliptical galaxies roughly consistent with that observed (e.g., Mamon 1992; Baugh, Cole \& Frenk 1996). An additional mechanism has been proposed by \citet{naab07}, but is mainly expected to be at work in the Early Universe forming primordial elliptical galaxies. 

Successions of several minor mergers are more likely than binary major mergers -- see \citet{ks06}, \citet{maller06}, and estimates in paper~I -- at least at redshift $z<1$. The number of elliptical galaxies is then a priori expected to increase significantly through this alternative mechanism. But then the number of elliptical galaxies present in the Universe today would be much larger than observed. Even among field spiral galaxies, which have significantly grown over the past 8 billion years, many would have become elliptical and lenticular galaxies. Another process is then required to prevent the succession of minor mergers from systematically transforming disk galaxies into ellipticals as soon as their mass is increased by a few ten of percents, otherwise this mechanism together with binary major mergers would result in an excess of elliptical galaxies. Still, there should be at least some ellipticals formed by multiple minor mergers at low redshift; \citet{vandokkum05} finds a large fraction of mass ratios larger than 4:1 in the mergers undergone recently by field ellipticals.

Observationally, on-going minor mergers are not uncommon at $z=0$, but also at $z\sim1$: for instance a large fraction of the interacting systems at $z\sim1$ in GEMS and GOODS is of M51-type \citep{elm07b}, which corresponds to mergers where the mass of the companion is far from negligible, as in our 15:1 -- 5:1 mergers. One merger of this type can transform a spiral into an Sa or S0 (paper~I) and a few ones make it elliptical. Galaxies thus have to accrete cold gas in a diffuse phase, or small clumps that are not too massive/concentrated so that they are disrupted before they reach the disk.

Nevertheless, multiple minor mergers must have formed some elliptical galaxies. In particular, some galaxies that are too massive to accrete cold gas \citep{birnboimdekel03} or in dense environments have likely evolved into ellipticals under the combined effect of repeated minor mergers (in addition to binary major mergers).
 This provides a good explanation for the boxiness of giant ellipticals, as discussed previously (Section~3.3), and there is further evidence that elliptical galaxies did not all form from binary major mergers \citep{naabostriker07}. Repeated major mergers in groups may have formed them too, but these are less likely than multiple minor mergers. Thus, successions of minor mergers likely participated in the formation of some of the elliptical galaxies seen today. 
 
Elliptical galaxies have a higher frequency of globular clusters (GCs) relative to their stellar mass than spiral galaxies \citep{harris91, vdb01}. The formation of GCs can be triggered by strong shocks associated with starbursts, as suggested by \citet{vdb79} and studied by \citet{jogsolomon92}. Major mergers trigger stronger starbursts, while the shocks and starbursts are less intense (but repeated) during the successive minor mergers (see Sect.~\ref{414}), potentially forming different numbers of GCs -- which models resolving GC formation could confirm. This could be a way to disentangle the elliptical galaxies formed by major and repeated minor mergers.


\section{Conclusion}

In this paper, we have shown that the succession of minor galaxy mergers (mass ratios larger than 4:1) leads to the gradual transformation of spiral galaxies into elliptical-like galaxies. This is the case both for the truly minor mergers (mass ratios above 10:1, a single merger of this type forms a spiral) and the ``intermediate'' mergers (mass ratios between 4:1 and 10:1, a single merger of this type forms an S0-like galaxy, e.g. paper~I). The remnants of repeated mergers with these mass ratios have a spheroidal shape, a radial profile close to the ``$R^{1/4}$'' empirical profile of ellipticals, and are supported by large velocity dispersions $V/\sigma$. This is true even for very low-mass companions (up to 50:1 in one of our simulations), and the relaxed remnant resembles an elliptical galaxy as soon as the total mass added by successive mergers exceeds 30--40\% of the mass of the main initial spiral. The global properties of the multiple merger remnants depend more on the total merged mass than on the mass ratio of each merger, and for instance a {\tt 10:10x1} merger sequence resembles a {\tt 1:1} remnant more than a {\tt 5:10x1}. These properties can vary with the orbital parameters and morphology of the progenitor galaxies, but not widely. In particular, only extremely gas-rich systems, or companions on peculiar orbits, may keep disk-dominated galaxies after several minor mergers, but this should not be the most frequent case except at very high redshift. A noteworthy dynamical result is that the final properties of the remnant (radial profile, flattening, kinematics) are to first order independent of whether it is formed by a single {\tt 1:1} merger, a  {\tt 4 x 4:1} sequence, or a  {\tt 10 x 10:1} sequence, so as long as the final mass has increased by the same factor.

Multiple, sequential minor mergers provide a new theoretical pathway to form elliptical galaxies without major mergers. In a {\emph{purely}} hierarchical scenario, a succession of several minor mergers is more likely than one single binary merger, and must then have formed some of the elliptical galaxies seen today. Repeated minor mergers also provide a possible explanation for some specific features, in particular the boxiness of the largest ellipticals, that are not explained by the standard major mergers (1:1-3:1).

However, minor mergers repeated in time would tend by themselves to produce too many elliptical galaxies. Moreover if field galaxies were to grow only by hierarchical merging, most spirals would have been transformed into ellipticals over the last 8 Gyr, even if these mergers were minor ones. It is then required that another process participates in the growth of galaxies, braking down or reversing the minor-merger-driven evolution towards early-type spheroids. This process is likely to be accretion of cold gas, either in a completely diffuse phase, or in small gas-rich clumps -- but not dwarf galaxies bounded by their dark matter haloes that would have destroyed massive disks.

While we present a new mechanism for the formation of elliptical galaxies in a purely hierarchical context, we also conclude that its efficiency must be somewhat limited: the evolution of galaxies cannot be purely merger-driven, and a large fraction of their mass cannot have been acquired through mergers, even minor ones with small companions. The growth of spiral galaxies at low and moderate redshifts must have rather been a competition between these minor mergers and accretion of cold gas. Some of the present-day ellipticals likely formed by multiple minor mergers, but this process cannot be much more efficient than binary major mergers, otherwise it would result in an excess of elliptical galaxies.

Cosmological simulations of restricted volumes can reach a resolution similar to that of the generic galaxy simulations presented in this paper \citep[see in particular][]{naab07}, but not over large samples that can be used for statistical purposes. At the opposite, large-volume cosmological simulations may lack the resolution needed to reproduce the effects of repeated minor mergers in a realistic way, because this requires that the dwarf companions are sufficiently resolved. However, these simulations can provide accurate predictions of the evolution history of galaxies (mass ratios of mergers, time intervals, and rate of diffuse gas accretion). The possibility of forming ellipticals but keeping most field galaxies as spirals at the same time, in high-resolution galaxy models taking into account these large-scale cosmological predictions, could be a future test of cosmological models.


\begin{acknowledgements}
We are very grateful to the anonymous referee whose critical comments helped us present the method and results clearly. Stimulating discussions on this work with Avishai~Dekel and Eric Emsellem and comments from Thorsten Naab on an earlier version of this manuscript are gratefully acknowledged. We are happy to acknowledge the support of the Indo-French grant IFCPAR/2704-1. The N-body simulations were computed on the NEC-SX8R of the CEA/CCRT computing center and the NEC-SX8 of CNRS/IDRIS. 
\end{acknowledgements}


\bibliographystyle{aa} 


\end{document}